\documentclass[sigconf]{acmart}
\pdfoutput=1
\usepackage{balance}
\usepackage{doi}

\def\tit{What Can (and Can't) we Do with Sparse Polynomials?}

\usepackage{multirow}
\usepackage[ruled,boxed,linesnumbered,vlined,noend]{algorithm2e}
\usepackage{etoolbox}

\usepackage{enumitem}
\usepackage[capitalise,noabbrev,nameinlink]{cleveref}
\Crefname{enumi}{Step}{Steps}
\usepackage{todonotes}

\newcommand{\ZZ}{\ensuremath{\mathbb{Z}}}
\newcommand{\NN}{\ensuremath{\mathbb{N}}}
\newcommand{\QQ}{\ensuremath{\mathbb{Q}}}

\newcommand{\F}{\ensuremath{\mathsf{F}}}
\newcommand{\R}{\ensuremath{\mathsf{R}}}
\newcommand{\GF}[1]{\ensuremath{\mathbb{F}_{#1}}}
\newcommand{\M}{\ensuremath{\mathsf{M}}}
\newcommand{\Poly}{\ensuremath{\mathbf{P}}}
\newcommand{\NP}{\ensuremath{\mathbf{NP}}}

\newcommand{\bnd}[2]{{\ensuremath{#1\mathopen{}\left(#2\right)\mathclose{}}}}
\newcommand{\oh}[1]{\bnd{O}{#1}}
\newcommand{\softoh}[1]{\bnd{\widetilde{O}}{#1}}

\newcommand{\height}[1]{\bnd{\mathsf{H}}{#1}}

\DeclareMathOperator{\llog}{loglog}

\theoremstyle{acmplain}
\newtheorem{theorem}{Theorem}

\theoremstyle{acmdefinition}
\newtheorem{openprob}[theorem]{Open Problem}

\newcommand{\eg}{\textit{e.g.}}
\newcommand{\ie}{\textit{i.e.}}

\DontPrintSemicolon

\setcopyright{usgov}

\acmDOI{10.1145/3208976.3209027}
\acmISBN{978-1-4503-5550-6/18/07}

\acmConference[ISSAC '18]{2018 ACM International Symposium on Symbolic and Algebraic Computation}{July 16--19, 2018}{New York, NY, USA}
\acmBooktitle{ISSAC '18: 2018 ACM International Symposium on Symbolic and Algebraic Computation, July 16--19, 2018, New York, NY, USA}
\acmYear{2018}
\copyrightyear{2018}
\startPage{25}

\begin{document}
\title{\tit}

\author{Daniel S.\ Roche}
\orcid{0000-0003-1408-6872}
\affiliation{%
  \institution{United States Naval Academy}
  \city{Annapolis}
  \state{Maryland}
  \country{U.S.A.}
}
\email{roche@usna.edu}

\begin{abstract}
  Simply put, a sparse polynomial is one whose zero coefficients are not explicitly
  stored. Such objects are ubiquitous in exact computing, and so
  naturally we would like to have efficient algorithms to handle them.
  However, with this compact storage comes new
  algorithmic challenges, as fast algorithms for dense polynomials may
  no longer be efficient. In this tutorial we examine the state of the
  art for sparse polynomial algorithms in three areas: arithmetic, interpolation,
  and factorization. The aim is to highlight recent progress both in
  theory and in practice, as well as opportunities for future work.
\end{abstract}


\maketitle

\section{Sparse Polynomials}

Sparse polynomials are found in the core of
nearly every computer algebra system or library, and polynomials with
many zero coefficients frequently occur in practical settings.

Mathematically, the dividing line between a sparse and dense polynomial
is not well-defined. From a computer science standpoint, there is a clear distinction,
depending on the \emph{representation} of that polynomial in memory: A
\emph{dense} representation stores zero coefficients explicitly and exponents
implicitly, whereas a \emph{sparse} representation does not store zero
coefficients at all, but stores exponents explicitly.

There are many variants of sparse representations \cite{Fat91}.
This tutorial considers algorithms for the most compact representation, the
so-called \emph{distributed sparse} storage \cite{MP11}. Let
$f\in\R[x_1,\ldots,x_n]$ be an $n$-variate polynomial with coefficients in a ring $\R$.
The representation of $f$ is by a list of $t$ nonzero terms
$$(c_1, e_{1,1},\ldots, e_{1,n}), (c_2, e_{2,1},\ldots, e_{2,n}),\ldots,
  (c_t, e_{t,1},\ldots, e_{t,n})
$$
such that
$$f = c_1 x_1^{e_{1,1}}\cdots x_n^{e_{1,n}}
  + c_2 x_1^{e_{2,1}}\cdots x_n^{e_{2,n}}
  + \cdots
  + c_t x_1^{e_{t,1}}\cdots x_n^{e_{t,n}}
$$
with each coefficient $c_i\in\R$ nonzero
and all exponent tuples $(e_{i,1},\ldots,e_{i,n})\in\NN^n$ distinct. We also
assume that the terms are sorted according to their exponents in some consistent way.

This sparse representation matches the one used by default for
multivariate polynomials in modern computer algebra systems and
libraries such as Magma \cite{magmapoly},
Maple \cite{MP13}, Mathematica, Sage \cite{sagepoly},
and Singular \cite{Sch03}.

\subsection{Sparse polynomial algorithm complexity}

Sparse polynomials in the distributed representation
are also called \emph{lacunary} or \emph{supersparse} \cite{KK05} in the
literature to emphasize that,
in this representation, the degree of a polynomial could be
exponentially larger than its bit-length. This
exposes the essential difficulty of computing with sparse polynomials,
in that \emph{efficient algorithms for dense polynomials may cost
exponential-time in the sparse setting}.

Specifically, when analyzing algorithms for dense polynomials, the most important
measure is the degree bound $D\in\NN$ such that $\deg f < D$. The size
of the dense representation of a univariate polynomial
is $D$ ring elements, and many operations
can be performed in $D^\oh{1}$ ring operations, or even $D(\log
D)^\oh{1}$.

In the sparse representation, we need to also consider the
number of nonzero terms $t$, and the \emph{bit-length} of the exponents.
For a multivariate (sparse) polynomial, the representation
size is $\oh{t}$ ring elements plus $\oh{nt\log D}$ bits, where $n$ is
the number of variables and $D$ is
now an upper bound on the maximum degree. The goal,
then, is to develop new sparse polynomial algorithms which minimize the
cost in terms of $n$, $t$, and $\log D$.

The coefficient ring $\R$ makes a difference for some
algorithms. In the general setting we let $\R$ be an arbitrary integral
domain, and count ring operations.
Another important setting is when $\R=\ZZ$,
the ring of integers, in which case we also account for
the size of coefficients. Write $\height{f}$ for the
\emph{height} of a polynomial, which is the maximum magnitude $\max_i
|c_i|$ of its coefficients; then a fast algorithm on $f$ should have a
small running time in terms of $\log \height{f}$.

For simplicity of presentation, and because the algorithmic work with
sparse polynomials is still at a much more coarse level than, say, that
of integers and dense polynomials, we frequently use the soft-oh
notation $\softoh{\gamma} := \oh{\gamma\cdot(\log\gamma)^{\oh{1}}}$,
where $\gamma$ is some running-time function.

\subsection{Overview}

The basic challenge of sparse polynomial algorithms is to match the
complexity of dense polynomial algorithms for the same task. In some
cases, interestingly, this is (provably) not possible --- for a few
problems, even a polynomial-time algorithm in the sparse representation
size would imply that $\Poly=\NP$.

Where algorithms are possible, one interesting feature is that they must
usually take into account not only the coefficient arithmetic over $\R$,
but also the \emph{exponent arithmetic} over $\ZZ$. In fact, the latter
frequently poses the most difficulty in the design of efficient
algorithms.

This tutorial aims to outline the state of sparse polynomial algorithms,
dividing roughly into the three areas of arithmetic, interpolation, and
factorization. We highlight where essentially-optimal algorithms are
already known, where they are known \emph{not} to exist, and the
numerous cases of open problems in between.

\section{Arithmetic}

Polynomials stored in the dense representation can be added and
subtracted in linear-time. Dense polynomial multiplication costs
$\oh{D^2}$ ring operations using the classical algorithm, but
considerable research effort has gone to reducing this complexity, which
we now denote as simply $\M(D)$.
The most general result of \cite{CK91} gives $\M(D)\in\oh{D\log
D\llog n}$, and more recent work \cite{Fur07,HHL17} reduces this even further
for most commonly-used rings.

Many other polynomial computations can be
reduced to multiplication. In particular, Euclidean division with
remainder costs $\oh{\M(D)}$, and (extended) gcd, multi-point
evaluation, interpolation, Chinese remaindering, and rational
reconstruction all cost $\oh{M(D)\log D}$ ring operations
\cite[\S 2--6]{BCG+17} \cite[\S 8--11]{vzGG03}. Note that all these
operations take quasi-linear time in the input size $\softoh{D}$.

As discussed previously,
these algorithms are \emph{not} polynomial-time in the size
of the sparse representation.

\subsection{Addition and subtraction}

Adding or subtracting sparse polynomials is a matter of combining like
terms, which amounts to a \emph{merge} operation on the two lists of
nonzero terms. From our assumption that terms are stored in sorted
order, this costs $\oh{t\log D}$ bit operations and $\oh{t}$ ring
additions, where $t$ is the number of terms in the two input
polynomials. This matches the size of the input and is therefore
optimal.

Notice, however, that the size of the \emph{output}
can grow much more quickly than with dense polynomials. When adding two
dense polynomials with degrees less than $D$, the cost is $\oh{D}$ and
the output size is also $D$; the size does not increase.
But when adding two $t$-sparse polynomials,
the number of nonzero terms in the output may double to $2t$. Hence
repeated addition of sparse polynomials (of which multiplication is a
special case) requires more care.

\subsection{Multiplication}

A significant difference from case of dense polynomial multiplication
is that the \emph{output size} grows quadratically: the product of
multiplying two $t$-sparse polynomials may have as many as $t^2$ nonzero
terms.

In terms of input size, therefore, the best that one can hope for is
$\oh{t^2}$ ring operations and $\oh{t^2\log D}$ bit complexity. This is
nearly achieved by the classical algorithm of repeated monomial
multiplications and additions, which has bit complexity $\oh{t^2\log
D\log t}$ if the additions (merges) are done in a balanced way
\cite{Joh74,Fat03,Fat08}.

In practice, the \emph{size of intermediate
results}, and in particular the time to handle memory allocation and
de-allocation, can dominate the complexity of multiplication. At least
one specialized data structure was designed for this purpose
\cite{Yan98}, and \cite{Joh74} suggested a very simple idea of using a
\emph{heap} to simultaneously store and sort un-merged terms in the
product.

Decades later, Monagan and Pearce \cite{MP07,MP09,MP13,MP15}
rediscovered the heaps idea and developed extremely efficient
implementations and a careful analysis. They point out that, while the
asymptotic runtime is the same, the space for intermediate results is
only $\oh{t}$, no matter how many terms are in the output. This has
tremendous practical performance benefits, and it seems that many
computer algebra systems now use this approach in at least some cases.
In fact, there seems to be considerable interest in the fastest
practical speeds for sparse polynomial multiplication on a variety of
hardware platforms including parallel computing
\cite{GL06,MP09,Bis12,GL13,PG16}.

Still, this is unsatisfying from an algorithmic perspective since the
number of ring and bit-operations is still quadratic in \emph{every}
case, meaning that even the heap-based algorithms will have a hard
cut-off with the speed of dense multiplication as polynomials start to
fill-in with nonzero coefficients.

Another approach is to consider the \emph{size of the output} as a
parameter in the complexity. In the worst case, this can be quadratic in
the input size, but in cases where the output is smaller, we can hope
for faster computation. Furthermore, considering the output size allows
for a smooth transition to dense algorithms, where the output is
guaranteed to have at most $2D$ nonzero terms.

\begin{openprob}\label{op:mult}
  Develop an algorithm to multiply two sparse polynomials $f,g\in\R[x]$
  using $\softoh{t\log D}$ ring and bit operations, where $t$ is the
  number of terms in $f$, $g$, and $fg$, and $D$ is an upper bound on
  their degrees.
\end{openprob}

Considerable progress has been made toward this open problem, and it
seems now \emph{nearly} within reach. Some authors have looked at
special cases, when the \emph{support} of nonzero coefficients has a
certain structure, to reduce to dense multiplication and achieve the
desired complexity in those cases \cite{Roc08,Roc11,HL12a}.

A more general solution approach is to use \emph{sparse interpolation}
algorithms, which we examine in further detail in the next section.
First, \cite{HL13} showed that the open problem is essentially solved
\emph{when the support of the product is already known}. More recently,
\cite{AR15,Arn16} solved \cref{op:mult} under two conditions:
\begin{itemize}
  \item The ring $\R$ is either the integers $\ZZ$, or some other ring
  such as $\GF{p}$ that can be reduced to that case with no loss of
  efficiency.
  \item The so-called \emph{structural support} matches the size of the
  \emph{actual support}. Which is to say (roughly) that there are not
  too many cancellations of coefficients in the product.
\end{itemize}
Removing the second condition seems to be the main remaining hurdle in solving
\cref{op:mult}.

\subsection{Division}

When dividing sparse polynomials,
it is \emph{imperative} to consider the
output size: for example, the exact division of two 2-term polynomials
$x^D-1$ by $x-1$ produces a quotient with $D$ nonzero terms.

Fortunately, the heaps idea which works well in practice for sparse
multiplication has also been adapted to sparse division, and this method
easily yields the remainder as well as the quotient
\cite{MP10,GL15}.
As before, this approach uses $\oh{t^2}$ ring operations, leaving us
with another challenge:

\begin{openprob}\label{op:div1}
  Given two sparse polynomials $f,g\in\R[x]$, develop an algorithm to
  compute the quotient and remainder $q,g\in\R[x]$ such that $f=qg+r$,
  using $\softoh{t\log D}$ ring and bit operations, where $t$ is the
  number of terms of $f$, $g$, $q$, and $g$, and $\deg f < D$.
\end{openprob}

Note that an efficient solution to \cref{op:div1} is already available
when $\deg g$ is small, \ie, when $f$ is sparse but $g$ is dense. The
algorithm in that case amounts to computing $x^{e_i}\bmod g$ for all
exponents $e_i$ that appear in the support of $f$, via repeated
squaring, then multiplying by the coefficients $c_i$ and summing.

In the more difficult case that $g$ is also sparse,
\cite[\S III]{DC09} has a nice discussion of the problem. In
particular, they point out that a seemingly-easier \emph{decision
problem} is still open:

\label{sec:divis}
\begin{openprob}\label{op:div2}
  Given two sparse polynomials $f,g\in\R[x]$, develop an algorithm which
  determines whether $g$ divides $f$ exactly, using $\softoh{t\log D}$
  ring and bit operations.
\end{openprob}

Again, a solution is known only when $g$ is dense.

\section{Interpolation}\label{sec:interp}

Polynomial interpolation is a problem of \emph{model
fitting}: given some measurements, find a (sparse) polynomial which
(best) fits the data.

In the case of dense polynomials, this is a classical problem.
Exact interpolation from an arbitrary set of points
can be accomplished in $\softoh{D}$ ring operations. Even if the data is
noisy or has outliers, classical numerical methods can recover the best
polynomial fit (see, \eg, \cite[\S 4,13]{Sol15}).

The challenge of \emph{sparse} polynomial interpolation is to
fit a $t$-sparse polynomial to some small number of evaluation points
$m\ll D$. Even the \emph{decidability} of this question is non-trivial
and depends on the choice of ring and evaluation points \cite{BT91}. In
fact, all efficient solutions require a stronger model that allows the
algorithm to sample the unknown function at arbitrary points.

\begin{definition}
  A \emph{black box} for an unknown $n$-variate polynomial $f\in\R[x_1,\ldots,x_n]$
  is a function which accepts any $n$-tuple
  $(\theta_1,\ldots,\theta_n)\in\R^n$ and produces the value
  $f(\theta_1,\ldots,\theta_n)\in\R$.
\end{definition}

In this context, a \emph{sparse interpolation algorithm} takes a black
box for an unknown $f$ as well as upper bounds $D,T$ on its degree and
number of nonzero terms, respectively. An efficient algorithm should
minimize the number of evaluations, plus required ring and
bit-operations, in terms of $T$ and $\log D$.

\begin{openprob}\label{op:interp}
  For any ring $\R$, given a black box for $f\in\R[x_1,\ldots,x_n]$ and
  bounds $T,D$ on the number of nonzero terms and maximum degree of $f$,
  determine the nonzero coefficients and corresponding exponents of $f$
  using $\softoh{T\log D}$ ring operations, bit operations, and
  black box evaluations.
\end{openprob}

The methods for sparse interpolation have very strong connections to
other techniques in coding theory and signal processing. In the first
case, the support of the unknown sparse polynomial $f$
corresponds to the error locations in Reed-Solomon decoding; the
decoding algorithm of Blahut \cite{Bla84} can be seen as a sparse
interpolation algorithm similar to Prony's method below \cite{CKP12}.

In another viewpoint, the evaluations of a sparse polynomial at integer
powers of some complex root of unity
have a 1-1 correspondence with evaluations of a sum of
exponentials at integer points: signal frequencies correspond to
polynomial exponents and amplitudes correspond to polynomial
coefficients. The recovery techniques come from multi-exponential
analysis, the theory of Pad\'e approximants, and tensor decomposition;
see \cite{CLW16,Mou17} and references therein.

In the context of computer algebra, algorithms for sparse polynomial
interpolation go back to the work of Zippel \cite{Zip79,Zip90}, who
developed a randomized algorithm which recovers a sparse polynomial in
recursively, variable-by variable. However, the reliance on dense
\emph{univariate} interpolation makes it unsuitable for the setting of
this tutorial.

For exact (super)sparse polynomial interpolation, we first consider the
easier case of univariate polynomials where $n=1$.
There are are essentially two classes of algorithms here, which we
discuss separately. Then we see how to reduce the multivariate case to
the univariate one without changing the sparsity.

\subsection{Prony's method}

The classic numerical technique of Prony \cite{Pro95} from the 18th
century was rediscovered by Ben-Or and Tiwari and adapted to the context of computer
algebra nearly 200 years later \cite{BT88}. The key idea is that any
sequence of evaluations at consecutive values in geometric progression
$(f(\omega^i))_{i\ge 0}$ form a linearly-recurrent sequence with degree $t$,
where $t$ is the actual number of nonzero terms in $f$. Furthermore,
the minimum polynomial of the linear recurrence is a product of linear
factors, and its roots are exactly of the form $\omega^{e_i}$, where
$e_i$ is the exponent of a nonzero term in $f$.

Given a black box for
unknown univariate sparse polynomial $f\in\R[z]$, plus degree and
sparsity bounds $D$ and $T$, the algorithm takes the following steps:
\begin{enumerate}
  \item Find a suitable element $\omega\in\R$ with multiplicative order
    at least $D$.
  \item Evaluate $f(1), f(\omega), f(\omega^2), \ldots,
    f(\omega^{2T-1})$.
  \item Use the Berlekamp-Massey algorithm (or a Toeplitz solver) to
    compute the minimum polynomial of the linear recurrence
    $\Lambda(z)\in\R[z]$.
  \item Find the roots $\omega^{e_i}$ of $\Lambda$.
  \item Compute the discrete logarithms of the roots to base $\omega$ to
    discover the exponents $e_1,\ldots,e_t$.
  \item Solve a transposed Vandermonde system from the first $t$
    evaluations to recover the nonzero coefficients $c_1,\ldots,c_t$.
\end{enumerate}

The main benefit of this procedure is that it computes the minimal
number of evaluations $2T$ in Step 2.
Steps 3 and 6 involve well-known techniques from structured linear
algebra and can be solved efficiently using $\softoh{t}$ ring operations
\cite{KY89,BLS03}.

However, the other steps depend heavily on the
coefficient ring $\R$. In particular, the algorithm must find a
high-order element $\omega\in\R$, perform root-finding of a degree-$t$ polynomial, and
then perform $t$ discrete logarithms to the base $\omega$.

When $\R=\ZZ$, these steps can be performed reasonably efficiently by
working modulo $p$ for a very carefully-chosen prime $p$.
The first modular version of this approach \cite{KLW90} was not
polynomial-time in the discrete logarithm step, but
an idea from \cite{Kal88,Kal10a} proposes choosing a prime $p$ with
$p \in \oh{D}$ such that $(p-1)$ is divisible by a large power of 2.
This divisor means that $\GF{p}^*$ has a large subgroup with smooth
order, facilitating fast discrete logarithms in only $\softoh{\log^2 D}$
field operations each. Because of the size of the prime, the resulting
algorithm has total cost of $\softoh{T\log^3 D + T\log\height{f}}$ bit
operations. This almost solves \cref{op:interp}, except it is cubic
in the size of the exponents $\log D$.

Many important improvements have been made to this algorithm since it
was developed in the 1990s. Early-termination techniques allow for only
$\oh{t}$ evaluations instead of $\oh{T}$, where $t\le T$ is the true
sparsity of $f$ \cite{KL03}. The root-finding step was found to be the
bottleneck in a practical implementation by \cite{HL15}; a better
root-finding algorithm in this case was developed \cite{GHL15} to
improve the situation.

\cite{JM10} adapt the algorithm to the case
of finite fields with a parallel algorithm that has better complexity in
terms of $\log D$ but becomes quadratic in the sparsity $t$; they also
implemented their algorithm and performed some experiments. Earlier
modular, parallel algorithms were also developed by \cite{GKS90,HR99}.
A more straightforward parallel algorithm over $\ZZ$ was developed and
experimentally evaluated by \cite{KRT15}.

\subsection{Homomorphic imaging}

As we have seen, the Prony approach to sparse interpolation does not
perform well over arbitrary finite fields due to the cost of discrete
logarithm computations, which in general cannot be performed in
polynomial-time.

A radically different method was first proposed by \cite{GS09}, based on
some earlier ideas of \cite{GK87}. This does not directly use the black
box model defined earlier, but instead assumes a more generous model that
can be stated as follows:

\begin{definition}
  A \emph{modular black box} for an unknown polynomial $f\in\R[x]$
  is a function which accepts any pair of dense polynomials
  $g,h\in\F[x]$ with $\deg h < \deg g$ and produces the value
  $f(h) \bmod g$.
\end{definition}

If $g=x$ and $h=\theta$, this corresponds to the normal black-box
evaluation $f(\theta)$. But when $\deg g > 1$, the setting is more
general. It makes sense when interpolating a straight-line program or
algebraic circuit for $f$, where each step of the evaluation can be
performed modulo $g$. We must be careful with the cost model also,
because for example if $\deg g > \deg f$, the problem is trivially solved
with a single evaluation. To accommodate this, we say that each such
evaluation costs $\softoh{\deg g}$ ring operations.

The first algorithm in this model by \cite{GS09} was deterministic and works over any
ring $\R$, but with a high complexity of $\softoh{T^4 \log^2 D}$. A
series of later improvements \cite{GR11a,AGR13,AGR14,AGR15,HG17}
has improved this to $\softoh{T\log^3 D}$ ring operations, largely by
introducing numerous randomizations. Note that this is a similar cost to
the best-known variants of Prony's method over $\ZZ$, but it still has
the comparative shortcoming of requiring more expensive
evaluations.

\subsection{Kronecker substitution}

Any $n$-variate polynomial $f$ with maximum degree less than $D$ is in
one-to-one correspondence with a univariate polynomial $g$ with degree less
than $D^n$, according to a map introduced by Kronecker \cite{Kro82}.
The forward map can be written as an evaluation of $f$ at powers of a
single variable $z$:
$$g(z) := f(z,z^D,\ldots,z^{D^{n-1}}).$$

The reverse map simply involves converting each integer exponent of $g$
into an $n$-tuple of exponents of $f$ via a $D$-adic expansion of the
univariate exponent.

Because the forward map is simply an evaluation, this means that a
multivariate $f$ can be found via univariate supersparse interpolation
of a single polynomial with degree less than $D^n$ and the same number
of nonzero terms. Supersparse algorithms have complexity in
terms of the bit-length of the exponents, so the resulting complexity
should be polynomial in $T$ and $n\log D$, as desired.

Even so, the exponential increase in degree is to be avoided, since the
cost of both approaches above is at least quadratic in $\log D$.
A compromise approach was presented by \cite{AR14},
who showed how to use a randomization to reduce the multivariate
polynomial to a set of $\oh{n}$ univariate polynomials, each of degree
only $\oh{DT}$. When combined with the univariate supersparse
interpolation algorithms above, this results in a better complexity in
terms of $n$.

\subsection{Problem variations and extensions}

Numerous authors have focused on solving different useful variants of the
sparse interpolation problem rather than improving the asymptotic
complexity. One important consideration is the \emph{basis}. So far we
have assumed a monomial basis $1,x,x^2,\ldots$, and the
\emph{arithmetic} algorithms of the previous section work more or less
the same over any basis. But interpolating into a different basis is
more subtle. Sparse interpolation in Pockhammer, Chebyshev, and shifted
power bases has been considered by \cite{LS95,GKL03,GR10,AK15,IKY18}.

Another interesting direction has been the development of more robust
sparse interpolation algorithms, which can tolerate numerical noise in
the evaluations, or completely-erroneous outliers, at the cost of
performing more evaluations than in the exact setting
\cite{KY07,BCK12,CKP12,AK15}.

An even more difficult problem is \emph{sparse rational function
interpolation}, which is the same as sparse polynomial interpolation
except that the unknown $f$ is a fraction of two sparse multivariate
polynomials. Interestingly, \cite{KN11} showed that the sparsest
rational function is not always reduced; see also
\cite{CL11,KY13,KPSW17}.

\section{Factorization}\label{sec:fact}

The development of efficient algorithms to factor dense polynomials has
been widely celebrated \cite{vzGP01,Kal03,HvHN11}. Most notably for our
current purposes, since
the 1980s it has been possible to factor polynomials over $\ZZ[x]$ in
polynomial-time. Ignoring the thorny issues with multivariate
polynomials and finite fields for now, we ask the same question for
sparse polynomials over $\ZZ[x]$.

This question is already addressed in some other surveys such as
\cite{Kal03,DC09,FS15}. Because of this, and since this is the area in
which the speaker has the least expertise, we give only a very cursory
overview of the accomplishments and challenges here.

\subsection{Impossibility results}

Plaisted \cite{Pla84} showed that the problem of determining whether two
sparse polynomials are relatively prime is $\NP$-complete, via a
reduction from 3-SAT. This means that even computing the gcd of two
supersparse polynomials is (seemingly) intractable.
However, as highlighted by \cite{DC09}, it is important to emphasize
that the reduction only uses cyclotomic polynomials; hence there is a
possibility that by excluding such polynomials more progress is
possible.

Another impossibility is complete factorization, as illustrated by
the (cyclotomic) example $x^D-1$, which has an exponentially-large dense
factor.

The best we can hope for is perhaps the following:
\begin{openprob}
  Suppose $f\in\ZZ[x]$ is a $t$-sparse polynomial with at least one
  sparse factor $g\in\ZZ[x]$ such that $g$ has at most $s$ nonzero
  terms. In polynomial-time in $t$, $s$, $\log\height{f}$, and
  $\log\deg f$, find any $s$-sparse factor of $f$.
\end{openprob}

\subsection{Low-degree factors}

One case in which supersparse polynomial factorization is possible is
when the factors are dense and have small degree. The results in this
category generally depend on \emph{gap lemmas}, whose statements are of
the following form: If $f\in\F[x]$ can be written as $f=f_0+f_1\cdot x^k$,
where the ``gap'' $(k-\deg f_0)$ is large, then every non-cyclotomic
factor of $f$ is a factor of both $f_0$ and $f_1$.

The actual gap lemmas are a bit more technical in specifying
the gap and some other conditions, but what they tell us is that finding
low-degree factors of a high-degree, sparse polynomial, can be reduced
to finding factors of some dense sub-polynomial(s) of $f$ and then
checking divisibility. (Recall from \cref{sec:divis} that sparse
divisibility testing is tractable when the divisor has low degree.)

This technique has been applied to degree-1 factors in $\ZZ[x]$
\cite{CKS99}, then to small degree over $\QQ[x]$ \cite{Len99}, degree-2 in
$\QQ[x,y]$ \cite{KK05}, and finally small degree over $\QQ[x_1,\ldots,x_n]$
\cite{KK06,Gre16a,CGKPS13j}.

\subsection{High-degree factors}

Finding high-degree sparse factors remains a challenge in almost all
cases. Very recent work by \cite{AS17} proves that essentially all
bivariate high-degree factors of a bivariate rational polynomial must be
sparse, which provides some new hope that this problem is tractable.

Otherwise, the only high-degree sparse factorizations that can be
computed efficiently are perfect roots of the form $f=g^k$ for some
$k\in\NN$. As shown by \cite{GR08,GR11}, such factors $g$ can be
computed when they exist and are sparse, and the power $k$ can be
computed unconditionally, in polynomial-time in the sparse size of $f$.
Interestingly, it can be proven that the power $k$ must be relatively
small whenever $f$ is sparse; conversely, a high power of any polynomial
is necessarily dense.

\begin{acks}
  This work was performed while the author was graciously 
  hosted by the Laboratoire
  Jean Kuntzmann at the Universit\'e Grenoble Alpes.

  This work was supported in part by the
  \grantsponsor{nsf}{National Science Foundation}{https://nsf.gov/}
  under grants
  \grantnum[https://www.nsf.gov/awardsearch/showAward?AWD_ID=1319994]{nsf}{1319994}
  and
  \grantnum[https://www.nsf.gov/awardsearch/showAward?AWD_ID=1618269]{nsf}{1618269}.
\end{acks}

\balance

\bibliographystyle{abbrvnat}

\begin{thebibliography}{87}
\providecommand{\natexlab}[1]{#1}
\providecommand{\url}[1]{\texttt{#1}}
\expandafter\ifx\csname urlstyle\endcsname\relax
  \providecommand{\doi}[1]{doi: #1}\else
  \providecommand{\doi}{doi: \begingroup \urlstyle{rm}\Url}\fi

\bibitem[Amoroso and Sombra(2017)]{AS17}
F.~Amoroso and M.~Sombra.
\newblock Factorization of bivariate sparse polynomials.
\newblock online, 2017.
\newblock URL \url{https://arxiv.org/abs/1710.11479}.

\bibitem[Arnold(2016)]{Arn16}
A.~Arnold.
\newblock \emph{Sparse Polynomial Interpolation and Testing}.
\newblock PhD thesis, University of Waterloo, 2016.
\newblock URL \url{http://hdl.handle.net/10012/10307}.

\bibitem[Arnold and Kaltofen(2015)]{AK15}
A.~Arnold and E.~L. Kaltofen.
\newblock Error-correcting sparse interpolation in the chebyshev basis.
\newblock ISSAC '15, pages 21--28. ACM, 2015.
\newblock \doi{10.1145/2755996.2756652}.

\bibitem[Arnold and Roche(2014)]{AR14}
A.~Arnold and D.~S. Roche.
\newblock Multivariate sparse interpolation using randomized {K}ronecker
  substitutions.
\newblock ISSAC '14, pages 35--42. ACM, 2014.
\newblock \doi{10.1145/2608628.2608674}.

\bibitem[Arnold and Roche(2015)]{AR15}
A.~Arnold and D.~S. Roche.
\newblock Output-sensitive algorithms for sumset and sparse polynomial
  multiplication.
\newblock ISSAC '15, pages 29--36. ACM, 2015.
\newblock \doi{10.1145/2755996.2756653}.

\bibitem[Arnold et~al.(2013)Arnold, Giesbrecht, and Roche]{AGR13}
A.~Arnold, M.~Giesbrecht, and D.~S. Roche.
\newblock Faster sparse interpolation of straight-line programs.
\newblock In V.~P. Gerdt, W.~Koepf, E.~W. Mayr, and E.~V. Vorozhtsov, editors,
  \emph{Proc. Computer Algebra in Scientific Computing (CASC 2013)}, volume
  8136 of \emph{Lecture Notes in Computer Science}, pages 61--74. Springer,
  September 2013.
\newblock \doi{10.1007/978-3-319-02297-0_5}.

\bibitem[Arnold et~al.(2014)Arnold, Giesbrecht, and Roche]{AGR14}
A.~Arnold, M.~Giesbrecht, and D.~S. Roche.
\newblock Sparse interpolation over finite fields via low-order roots of unity.
\newblock ISSAC '14, pages 27--34. ACM, 2014.
\newblock \doi{10.1145/2608628.2608671}.

\bibitem[Arnold et~al.(2015)Arnold, Giesbrecht, and Roche]{AGR15}
A.~Arnold, M.~Giesbrecht, and D.~S. Roche.
\newblock Faster sparse multivariate polynomial interpolation of straight-line
  programs.
\newblock \emph{Journal of Symbolic Computation}, 2015.
\newblock \doi{10.1016/j.jsc.2015.11.005}.

\bibitem[Ben-Or and Tiwari(1988)]{BT88}
M.~Ben-Or and P.~Tiwari.
\newblock A deterministic algorithm for sparse multivariate polynomial
  interpolation.
\newblock STOC '88, pages 301--309. ACM, 1988.
\newblock \doi{10.1145/62212.62241}.

\bibitem[Biscani(2012)]{Bis12}
F.~Biscani.
\newblock Parallel sparse polynomial multiplication on modern hardware
  architectures.
\newblock ISSAC '12, 2012.

\bibitem[Blahut(1984)]{Bla84}
R.~E. Blahut.
\newblock A universal reed-solomon decoder.
\newblock \emph{IBM Journal of Research and Development}, 28\penalty0
  (2):\penalty0 150--158, March 1984.
\newblock \doi{10.1147/rd.282.0150}.

\bibitem[Borodin and Tiwari(1991)]{BT91}
A.~Borodin and P.~Tiwari.
\newblock On the decidability of sparse univariate polynomial interpolation.
\newblock \emph{Computational Complexity}, 1:\penalty0 67--90, 1991.
\newblock \doi{10.1007/BF01200058}.

\bibitem[Bostan et~al.(2003)Bostan, Lecerf, and Schost]{BLS03}
A.~Bostan, G.~Lecerf, and E.~Schost.
\newblock Tellegen's principle into practice.
\newblock ISSAC '03, pages 37--44. ACM, 2003.
\newblock \doi{10.1145/860854.860870}.

\bibitem[Bostan et~al.(2017)Bostan, Chyzak, Giusti, Lebreton, Lecerf, Salvy,
  and Schost]{BCG+17}
A.~Bostan, F.~Chyzak, M.~Giusti, R.~Lebreton, G.~Lecerf, B.~Salvy, and
  E.~Schost.
\newblock \emph{Algorithmes Efficaces en Calcul Formel}.
\newblock 1.0 edition, Aug. 2017.

\bibitem[Boyer et~al.(2012)Boyer, Comer, and Kaltofen]{BCK12}
B.~Boyer, M.~T. Comer, and E.~L. Kaltofen.
\newblock Sparse polynomial interpolation by variable shift in the presence of
  noise and outliers in the evaluations.
\newblock In \emph{Electr. Proc. Tenth Asian Symposium on Computer Mathematics
  (ASCM 2012)}, 2012.

\bibitem[Cantor and Kaltofen(1991)]{CK91}
D.~G. Cantor and E.~Kaltofen.
\newblock On fast multiplication of polynomials over arbitrary algebras.
\newblock \emph{Acta Informatica}, 28:\penalty0 693--701, 1991.
\newblock \doi{10.1007/BF01178683}.

\bibitem[Chattopadhyay et~al.(2013)Chattopadhyay, Grenet, Koiran, Portier, and
  Strozecki]{CGKPS13j}
A.~Chattopadhyay, B.~Grenet, P.~Koiran, N.~Portier, and Y.~Strozecki.
\newblock {Computing the multilinear factors of lacunary polynomials without
  heights}.
\newblock Manuscript (submitted), 2013.
\newblock URL \url{https://arxiv.org/abs/1311.5694}.

\bibitem[Comer et~al.(2012)Comer, Kaltofen, and Pernet]{CKP12}
M.~T. Comer, E.~L. Kaltofen, and C.~Pernet.
\newblock Sparse polynomial interpolation and {Berlekamp/Massey} algorithms
  that correct outlier errors in input values.
\newblock ISSAC '12, pages 138--145. ACM, 2012.
\newblock \doi{10.1145/2442829.2442852}.

\bibitem[Cucker et~al.(1999)Cucker, Koiran, and Smale]{CKS99}
F.~Cucker, P.~Koiran, and S.~Smale.
\newblock A polynomial time algorithm for {D}iophantine equations in one
  variable.
\newblock \emph{J. Symbolic Comput.}, 27\penalty0 (1):\penalty0 21--29, 1999.
\newblock \doi{10.1006/jsco.1998.0242}.

\bibitem[Cuyt and Lee(2011)]{CL11}
A.~Cuyt and W.~Lee.
\newblock Sparse interpolation of multivariate rational functions.
\newblock \emph{Theoretical Computer Science}, 412\penalty0 (16):\penalty0 1445
  -- 1456, 2011.
\newblock \doi{10.1016/j.tcs.2010.11.050}.

\bibitem[Cuyt et~al.(2016)Cuyt, shin Lee, and Wang]{CLW16}
A.~Cuyt, W.~shin Lee, and X.~Wang.
\newblock On tensor decomposition, sparse interpolation and {Pad\'e}
  approximation.
\newblock \emph{Jaen journal on approximation}, 8\penalty0 (1):\penalty0
  33--58, 2016.

\bibitem[Davenport and Carette(2009)]{DC09}
J.~H. Davenport and J.~Carette.
\newblock The sparsity challenges.
\newblock In \emph{Symbolic and Numeric Algorithms for Scientific Computing
  (SYNASC), 2009 11th International Symposium on}, pages 3 --7, Sept. 2009.
\newblock \doi{10.1109/SYNASC.2009.62}.

\bibitem[Fateman(2003)]{Fat03}
R.~Fateman.
\newblock Comparing the speed of programs for sparse polynomial multiplication.
\newblock \emph{SIGSAM Bull.}, 37\penalty0 (1):\penalty0 4--15, March 2003.
\newblock \doi{10.1145/844076.844080}.

\bibitem[Fateman(2008)]{Fat08}
R.~Fateman.
\newblock Draft: {W}hat's it worth to write a short program for polynomial
  multiplication?
\newblock Online, Dec. 2008.
\newblock URL \url{http://www.cs.berkeley.edu/~fateman/papers/shortprog.pdf}.

\bibitem[Fateman(1991)]{Fat91}
R.~J. Fateman.
\newblock Endpaper: Frpoly: A benchmark revisited.
\newblock \emph{LISP and Symbolic Computation}, 4\penalty0 (2):\penalty0
  155--164, Apr 1991.
\newblock \doi{10.1007/BF01813018}.

\bibitem[Forbes and Shpilka(2015)]{FS15}
M.~A. Forbes and A.~Shpilka.
\newblock Complexity theory column 88: Challenges in polynomial factorization.
\newblock \emph{SIGACT News}, 46\penalty0 (4):\penalty0 32--49, Dec. 2015.
\newblock \doi{10.1145/2852040.2852051}.

\bibitem[F{\"u}rer(2007)]{Fur07}
M.~F{\"u}rer.
\newblock Faster integer multiplication.
\newblock STOC '07, pages 57--66. ACM, 2007.
\newblock \doi{10.1145/1250790.1250800}.

\bibitem[Garg and Schost(2009)]{GS09}
S.~Garg and {\'E}.~Schost.
\newblock Interpolation of polynomials given by straight-line programs.
\newblock \emph{Theoretical Computer Science}, 410\penalty0 (27-29):\penalty0
  2659--2662, 2009.
\newblock \doi{10.1016/j.tcs.2009.03.030}.

\bibitem[Gastineau and Laskar(2006)]{GL06}
M.~Gastineau and J.~Laskar.
\newblock Development of {TRIP}: Fast sparse multivariate polynomial
  multiplication using burst tries.
\newblock In V.~Alexandrov, G.~van Albada, P.~Sloot, and J.~Dongarra, editors,
  \emph{Computational Science - ICCS 2006}, volume 3992 of \emph{Lecture Notes
  in Computer Science}, pages 446--453. Springer Berlin Heidelberg, 2006.
\newblock \doi{10.1007/11758525_60}.

\bibitem[Gastineau and Laskar(2013)]{GL13}
M.~Gastineau and J.~Laskar.
\newblock Highly scalable multiplication for distributed sparse multivariate
  polynomials on many-core systems.
\newblock In V.~P. Gerdt, W.~Koepf, E.~W. Mayr, and E.~V. Vorozhtsov, editors,
  \emph{Computer Algebra in Scientific Computing}, pages 100--115, Cham, 2013.
  Springer International Publishing.
\newblock \doi{10.1007/978-3-319-02297-0_8}.

\bibitem[Gastineau and Laskar(2015)]{GL15}
M.~Gastineau and J.~Laskar.
\newblock Parallel sparse multivariate polynomial division.
\newblock PASCO '15, pages 25--33. ACM, 2015.
\newblock \doi{10.1145/2790282.2790285}.

\bibitem[\Gathen{von zur Gathen} and Gerhard(2003)]{vzGG03}
J.~\Gathen{von zur Gathen} and J.~Gerhard.
\newblock \emph{Modern Computer Algebra}.
\newblock Cambridge University Press, Cambridge, second edition, 2003.

\bibitem[\Gathen{von zur Gathen} and Panario(2001)]{vzGP01}
J.~\Gathen{von zur Gathen} and D.~Panario.
\newblock Factoring polynomials over finite fields: A survey.
\newblock \emph{Journal of Symbolic Computation}, 31\penalty0 (1-2):\penalty0 3
  -- 17, 2001.
\newblock \doi{10.1006/jsco.1999.1002}.

\bibitem[Giesbrecht and Roche(2008)]{GR08}
M.~Giesbrecht and D.~S. Roche.
\newblock On lacunary polynomial perfect powers.
\newblock ISSAC '08, pages 103--110. ACM, 2008.
\newblock \doi{10.1145/1390768.1390785}.

\bibitem[Giesbrecht and Roche(2010)]{GR10}
M.~Giesbrecht and D.~S. Roche.
\newblock Interpolation of shifted-lacunary polynomials.
\newblock \emph{Computational Complexity}, 19:\penalty0 333--354, 2010.
\newblock \doi{10.1007/s00037-010-0294-0}.

\bibitem[Giesbrecht and Roche(2011{\natexlab{a}})]{GR11}
M.~Giesbrecht and D.~S. Roche.
\newblock Detecting lacunary perfect powers and computing their roots.
\newblock \emph{Journal of Symbolic Computation}, 46\penalty0 (11):\penalty0
  1242--1259, 2011{\natexlab{a}}.
\newblock \doi{10.1016/j.jsc.2011.08.006}.

\bibitem[Giesbrecht and Roche(2011{\natexlab{b}})]{GR11a}
M.~Giesbrecht and D.~S. Roche.
\newblock Diversification improves interpolation.
\newblock ISSAC '11, pages 123--130. ACM, 2011{\natexlab{b}}.
\newblock \doi{10.1145/1993886.1993909}.

\bibitem[Giesbrecht et~al.(2003)Giesbrecht, Kaltofen, and Lee]{GKL03}
M.~Giesbrecht, E.~Kaltofen, and W.~Lee.
\newblock Algorithms for computing sparsest shifts of polynomials in power,
  chebyshev, and pochhammer bases.
\newblock \emph{Journal of Symbolic Computation}, 36\penalty0 (3-4):\penalty0
  401 -- 424, 2003.
\newblock \doi{10.1016/S0747-7171(03)00087-7}.
\newblock ISSAC 2002.

\bibitem[Grenet(2016)]{Gre16a}
B.~Grenet.
\newblock Bounded-degree factors of lacunary multivariate polynomials.
\newblock \emph{Journal of Symbolic Computation}, 75:\penalty0 171--192, 2016.
\newblock \doi{10.1016/j.jsc.2015.11.013}.
\newblock Special issue on the conference ISSAC 2014: Symbolic computation and
  computer algebra.

\bibitem[Grenet et~al.(2015)Grenet, \Hoeven{van der Hoeven}, and Lecerf]{GHL15}
B.~Grenet, J.~\Hoeven{van der Hoeven}, and G.~Lecerf.
\newblock Randomized root finding over finite {FFT}-fields using tangent
  {G}raeffe transforms.
\newblock In \emph{Proc. 40th International Symposium on Symbolic and Algebraic
  Computation}, ISSAC '15, page to appear, 2015.

\bibitem[Grigoriev and Karpinski(1987)]{GK87}
D.~Y. Grigoriev and M.~Karpinski.
\newblock The matching problem for bipartite graphs with polynomially bounded
  permanents is in {NC}.
\newblock In \emph{Foundations of Computer Science, 1987., 28th Annual
  Symposium on}, pages 166--172, Oct. 1987.
\newblock \doi{10.1109/SFCS.1987.56}.

\bibitem[Grigoriev et~al.(1990)Grigoriev, Karpinski, and Singer]{GKS90}
D.~Y. Grigoriev, M.~Karpinski, and M.~F. Singer.
\newblock Fast parallel algorithms for sparse multivariate polynomial
  interpolation over finite fields.
\newblock \emph{SIAM Journal on Computing}, 19\penalty0 (6):\penalty0
  1059--1063, 1990.
\newblock \doi{10.1137/0219073}.

\bibitem[Hart et~al.(2011)Hart, van Hoeij, and Novocin]{HvHN11}
W.~Hart, M.~van Hoeij, and A.~Novocin.
\newblock Practical polynomial factoring in polynomial time.
\newblock ISSAC '11, pages 163--170. ACM, 2011.
\newblock \doi{10.1145/1993886.1993914}.

\bibitem[Harvey et~al.(2017)Harvey, \Hoeven{van der Hoeven}, and Lecerf]{HHL17}
D.~Harvey, J.~\Hoeven{van der Hoeven}, and G.~Lecerf.
\newblock Faster polynomial multiplication over finite fields.
\newblock \emph{J. ACM}, 63\penalty0 (6):\penalty0 52:1--52:23, Jan. 2017.
\newblock \doi{10.1145/3005344}.

\bibitem[\Hoeven{van der Hoeven} and Lecerf(2012)]{HL12a}
J.~\Hoeven{van der Hoeven} and G.~Lecerf.
\newblock On the complexity of multivariate blockwise polynomial
  multiplication.
\newblock In \emph{Proc. ISSAC 2012}, pages 211--218, 2012.
\newblock \doi{10.1145/2442829.2442861}.

\bibitem[\Hoeven{van der Hoeven} and Lecerf(2013)]{HL13}
J.~\Hoeven{van der Hoeven} and G.~Lecerf.
\newblock On the bit-complexity of sparse polynomial and series multiplication.
\newblock \emph{Journal of Symbolic Computation}, 50:\penalty0 227--0254, 2013.
\newblock \doi{10.1016/j.jsc.2012.06.004}.

\bibitem[\Hoeven{van der Hoeven} and Lecerf(2015)]{HL15}
J.~\Hoeven{van der Hoeven} and G.~Lecerf.
\newblock Sparse polynomial interpolation in practice.
\newblock \emph{ACM Commun. Comput. Algebra}, 48\penalty0 (3/4):\penalty0
  187--191, Feb. 2015.
\newblock \doi{10.1145/2733693.2733721}.

\bibitem[Huang and Rao(1999)]{HR99}
M.-D.~A. Huang and A.~J. Rao.
\newblock Interpolation of sparse multivariate polynomials over large finite
  fields with applications.
\newblock \emph{Journal of Algorithms}, 33\penalty0 (2):\penalty0 204--228,
  1999.
\newblock \doi{10.1006/jagm.1999.1045}.

\bibitem[Huang and Gao(2017)]{HG17}
Q.~Huang and X.~Gao.
\newblock Faster deterministic sparse interpolation algorithms for
  straight-line program multivariate polynomials.
\newblock \emph{CoRR}, abs/1709.08979, 2017.
\newblock URL \url{http://arxiv.org/abs/1709.08979}.

\bibitem[Imamoglu et~al.(2018)Imamoglu, Kaltofen, and Yang]{IKY18}
E.~Imamoglu, E.~L. Kaltofen, and Z.~Yang.
\newblock Sparse polynomial interpolation with arbitrary orthogonal polynomial
  bases.
\newblock In \emph{Proc. ISSAC'18}, 2018.

\bibitem[Javadi and Monagan(2010)]{JM10}
S.~M.~M. Javadi and M.~Monagan.
\newblock Parallel sparse polynomial interpolation over finite fields.
\newblock PASCO '10, pages 160--168. ACM, 2010.
\newblock \doi{10.1145/1837210.1837233}.

\bibitem[Johnson(1974)]{Joh74}
S.~C. Johnson.
\newblock Sparse polynomial arithmetic.
\newblock \emph{SIGSAM Bull.}, 8:\penalty0 63--71, August 1974.
\newblock \doi{10.1145/1086837.1086847}.

\bibitem[Kaltofen(1988)]{Kal88}
E.~Kaltofen.
\newblock Notes on polynomial and rational function interpolation.
\newblock Unpublished manuscript, 1988.

\bibitem[Kaltofen(2003)]{Kal03}
E.~Kaltofen.
\newblock Polynomial factorization: A success story.
\newblock ISSAC '03, pages 3--4. ACM, 2003.
\newblock \doi{10.1145/860854.860857}.

\bibitem[Kaltofen and Koiran(2005)]{KK05}
E.~Kaltofen and P.~Koiran.
\newblock On the complexity of factoring bivariate supersparse (lacunary)
  polynomials.
\newblock ISSAC '05, pages 208--215. ACM, 2005.
\newblock \doi{10.1145/1073884.1073914}.

\bibitem[Kaltofen and Koiran(2006)]{KK06}
E.~Kaltofen and P.~Koiran.
\newblock Finding small degree factors of multivariate supersparse (lacunary)
  polynomials over algebraic number fields.
\newblock ISSAC '06, pages 162--168. ACM, 2006.
\newblock \doi{10.1145/1145768.1145798}.

\bibitem[Kaltofen and Lee(2003)]{KL03}
E.~Kaltofen and W.~Lee.
\newblock Early termination in sparse interpolation algorithms.
\newblock \emph{Journal of Symbolic Computation}, 36\penalty0 (3-4):\penalty0
  365--400, 2003.
\newblock \doi{10.1016/S0747-7171(03)00088-9}.
\newblock ISSAC 2002.

\bibitem[Kaltofen and Yagati(1989)]{KY89}
E.~Kaltofen and L.~Yagati.
\newblock Improved sparse multivariate polynomial interpolation algorithms.
\newblock In P.~Gianni, editor, \emph{Symbolic and Algebraic Computation},
  volume 358 of \emph{Lecture Notes in Computer Science}, pages 467--474.
  Springer Berlin / Heidelberg, 1989.
\newblock \doi{10.1007/3-540-51084-2_44}.

\bibitem[Kaltofen and Yang(2007)]{KY07}
E.~Kaltofen and Z.~Yang.
\newblock On exact and approximate interpolation of sparse rational functions.
\newblock ISSAC '07, pages 203--210. ACM, 2007.
\newblock \doi{10.1145/1277548.1277577}.

\bibitem[Kaltofen et~al.(1990)Kaltofen, Lakshman, and Wiley]{KLW90}
E.~Kaltofen, Y.~N. Lakshman, and J.-M. Wiley.
\newblock Modular rational sparse multivariate polynomial interpolation.
\newblock ISSAC '90, pages 135--139. ACM, 1990.
\newblock \doi{10.1145/96877.96912}.

\bibitem[Kaltofen(2010)]{Kal10a}
E.~L. Kaltofen.
\newblock Fifteen years after {DSC} and {WLSS2}: {W}hat parallel computations
  {I} do today [invited lecture at {PASCO} 2010].
\newblock PASCO '10, pages 10--17. ACM, 2010.
\newblock \doi{10.1145/1837210.1837213}.

\bibitem[Kaltofen and Nehring(2011)]{KN11}
E.~L. Kaltofen and M.~Nehring.
\newblock Supersparse black box rational function interpolation.
\newblock ISSAC '11, pages 177--186. ACM, 2011.
\newblock \doi{10.1145/1993886.1993916}.

\bibitem[Kaltofen and Yang(2013)]{KY13}
E.~L. Kaltofen and Z.~Yang.
\newblock Sparse multivariate function recovery from values with noise and
  outlier errors.
\newblock ISSAC '13, pages 219--226. ACM, 2013.
\newblock \doi{10.1145/2465506.2465524}.

\bibitem[Kaltofen et~al.(2017)Kaltofen, Pernet, Storjohann, and
  Waddell]{KPSW17}
E.~L. Kaltofen, C.~Pernet, A.~Storjohann, and C.~Waddell.
\newblock Early termination in parametric linear system solving and rational
  function vector recovery with error correction.
\newblock ISSAC '17, pages 237--244. ACM, 2017.
\newblock \doi{10.1145/3087604.3087645}.

\bibitem[Khochtali et~al.(2015)Khochtali, Roche, and Tian]{KRT15}
M.~Khochtali, D.~S. Roche, and X.~Tian.
\newblock Parallel sparse interpolation using small primes.
\newblock {PASCO} '15, pages 70--77. ACM, 2015.
\newblock \doi{10.1145/2790282.2790290}.

\bibitem[Kronecker(1882)]{Kro82}
L.~Kronecker.
\newblock {Grundz{\"u}ge einer arithmetischen Theorie der algebraischen
  Gr{\"o}ssen}.
\newblock \emph{Journal f{\"u}r die reine und angewandte Mathematik},
  92:\penalty0 1--122, 1882.

\bibitem[Lakshman and Saunders(1995)]{LS95}
Y.~N. Lakshman and B.~D. Saunders.
\newblock Sparse polynomial interpolation in nonstandard bases.
\newblock \emph{SIAM Journal on Computing}, 24\penalty0 (2):\penalty0 387--397,
  1995.
\newblock \doi{10.1137/S0097539792237784}.

\bibitem[Lenstra(1999)]{Len99}
H.~W. Lenstra, Jr.
\newblock Finding small degree factors of lacunary polynomials.
\newblock In \emph{Number theory in progress, {V}ol. 1
  ({Z}akopane-{K}o\'scielisko, 1997)}, pages 267--276. de Gruyter, Berlin,
  1999.

\bibitem[Monagan and Pearce(2007)]{MP07}
M.~Monagan and R.~Pearce.
\newblock Polynomial division using dynamic arrays, heaps, and packed exponent
  vectors.
\newblock In V.~Ganzha, E.~Mayr, and E.~Vorozhtsov, editors, \emph{Computer
  Algebra in Scientific Computing}, volume 4770 of \emph{Lecture Notes in
  Computer Science}, pages 295--315. Springer Berlin / Heidelberg, 2007.
\newblock \doi{10.1007/978-3-540-75187-8_23}.

\bibitem[Monagan and Pearce(2009)]{MP09}
M.~Monagan and R.~Pearce.
\newblock Parallel sparse polynomial multiplication using heaps.
\newblock ISSAC '09, pages 263--270. ACM, 2009.
\newblock \doi{10.1145/1576702.1576739}.

\bibitem[Monagan and Pearce(2010)]{MP10}
M.~Monagan and R.~Pearce.
\newblock Sparse polynomial division using a heap.
\newblock \emph{Journal of Symbolic Computation}, In Press, Corrected Proof,
  2010.
\newblock \doi{10.1016/j.jsc.2010.08.014}.

\bibitem[Monagan and Pearce(2011)]{MP11}
M.~Monagan and R.~Pearce.
\newblock Sparse polynomial multiplication and division in maple 14.
\newblock \emph{ACM Commun. Comput. Algebra}, 44\penalty0 (3/4):\penalty0
  205--209, Jan. 2011.
\newblock \doi{10.1145/1940475.1940521}.

\bibitem[Monagan and Pearce(2013)]{MP13}
M.~Monagan and R.~Pearce.
\newblock {POLY}: A new polynomial data structure for maple 17.
\newblock \emph{ACM Commun. Comput. Algebra}, 46\penalty0 (3/4):\penalty0
  164--167, Jan. 2013.
\newblock \doi{10.1145/2429135.2429173}.

\bibitem[Monagan and Pearce(2015)]{MP15}
M.~Monagan and R.~Pearce.
\newblock The design of {Maple}'s sum-of-products and {POLY} data structures
  for representing mathematical objects.
\newblock \emph{ACM Commun. Comput. Algebra}, 48\penalty0 (3/4):\penalty0
  166--186, Feb. 2015.
\newblock \doi{10.1145/2733693.2733720}.

\bibitem[Mourrain(2017)]{Mou17}
B.~Mourrain.
\newblock Fast algorithm for border bases of artinian gorenstein algebras.
\newblock ISSAC '17, pages 333--340. ACM, 2017.
\newblock \doi{10.1145/3087604.3087632}.

\bibitem[Plaisted(1984)]{Pla84}
D.~A. Plaisted.
\newblock New {NP}-hard and {NP}-complete polynomial and integer divisibility
  problems.
\newblock \emph{Theoret. Comput. Sci.}, 31\penalty0 (1-2):\penalty0 125--138,
  1984.
\newblock \doi{10.1016/0304-3975(84)90130-0}.

\bibitem[Popescu and Garcia(2016)]{PG16}
D.~A. Popescu and R.~T. Garcia.
\newblock Multivariate polynomial multiplication on gpu.
\newblock \emph{Procedia Computer Science}, 80:\penalty0 154 -- 165, 2016.
\newblock \doi{10.1016/j.procs.2016.05.306}.
\newblock International Conference on Computational Science 2016, ICCS 2016,
  6-8 June 2016, San Diego, California, USA.

\bibitem[Prony(1795)]{Pro95}
B.~d. Prony.
\newblock Essai exp\'erimental et analytique sur les lois de la
  {D}ilatabilit\'e des fluides \'elastique et sur celles de la {F}orce
  expansive de la vapeur de l’eau et de la vapeur de l’alkool, \`a
  diff\'erentes temp\'eratures.
\newblock \emph{J. de l’\'Ecole Polytechnique}, 1:\penalty0 24--76, 1795.

\bibitem[Roche(2008)]{Roc08}
D.~S. Roche.
\newblock Adaptive polynomial multiplication.
\newblock In \emph{Proc. Milestones in Computer Algebra (MICA)}, pages 65--72,
  2008.

\bibitem[Roche(2011)]{Roc11}
D.~S. Roche.
\newblock Chunky and equal-spaced polynomial multiplication.
\newblock \emph{Journal of Symbolic Computation}, 46\penalty0 (7):\penalty0
  791--806, July 2011.
\newblock \doi{10.1016/j.jsc.2010.08.013}.

\bibitem[Sch{\"o}nemann(2003)]{Sch03}
H.~Sch{\"o}nemann.
\newblock Singular in a framework for polynomial computations.
\newblock In M.~Joswig and N.~Takayama, editors, \emph{Algebra, Geometry and
  Software Systems}, pages 163--176, Berlin, Heidelberg, 2003. Springer Berlin
  Heidelberg.
\newblock \doi{10.1007/978-3-662-05148-1_9}.

\bibitem[Solomon(2015)]{Sol15}
J.~Solomon.
\newblock \emph{Numerical Algorithms}.
\newblock AK Peters/CRC Press, 2015.

\bibitem[Steel(2018)]{magmapoly}
A.~Steel.
\newblock Multivariate polynomial rings.
\newblock In \emph{The {Magma} Handbook}. Computational Algebra Group,
  University of Sydney, 2018.
\newblock URL
  \url{http://magma.maths.usyd.edu.au/magma/handbook/text/223#1924}.

\bibitem[Stein and Sage Development~Team()]{sagepoly}
W.~Stein and T.~Sage Development~Team.
\newblock Polynomial rings.
\newblock In \emph{Sage Reference Manual}.
\newblock URL
  \url{https://doc.sagemath.org/html/en/reference/polynomial_rings}.
\newblock v8.2.

\bibitem[Yan(1998)]{Yan98}
T.~Yan.
\newblock The geobucket data structure for polynomials.
\newblock \emph{Journal of Symbolic Computation}, 25\penalty0 (3):\penalty0
  285--293, 1998.
\newblock \doi{10.1006/jsco.1997.0176}.

\bibitem[Zippel(1979)]{Zip79}
R.~Zippel.
\newblock Probabilistic algorithms for sparse polynomials.
\newblock In E.~Ng, editor, \emph{Symbolic and Algebraic Computation},
  volume~72 of \emph{Lecture Notes in Computer Science}, pages 216--226.
  Springer Berlin / Heidelberg, 1979.
\newblock \doi{10.1007/3-540-09519-5_73}.

\bibitem[Zippel(1990)]{Zip90}
R.~Zippel.
\newblock Interpolating polynomials from their values.
\newblock \emph{Journal of Symbolic Computation}, 9\penalty0 (3):\penalty0
  375--403, 1990.
\newblock \doi{10.1016/S0747-7171(08)80018-1}.
\newblock Computational algebraic complexity editorial.

\end{thebibliography}

\newcommand{\Gathen}{\relax}\newcommand{\Hoeven}{\relax}

\end{document}